\RequirePackage{fix-cm}
\documentclass[smallextended]{svjour3}       
\smartqed  
\usepackage{amsmath}
\usepackage{graphicx}
\usepackage{subcaption}
\usepackage{algorithmic}
\usepackage{url}
\usepackage{algorithm2e}

 \journalname{Networks and Spatial Economics}
\begin{document}

\title{World City Networks and Multinational Firms: An Analysis of Economic Ties Over a Decade}


\titlerunning{Analyzing Topological Overlap of Cities}        
\author{Mohammed Adil Saleem \and Faraz Zaidi\footnote{Corresponding author: Faraz Zaidi, \email{faraz.a.zaidi@ieee.org}} \and C\'{e}line Rozenblat 
}

\institute{Saleem, MA \at School of Mathematics and Computer Science, Institute of Business Administration, Karachi, Pakistan.
\email{adilsaleem@iba.edu.pk}
\and
Zaidi, F \at School of Administrative Studies, York University, Canada and School of Mathematics and Computer Science, Institute of Business Administration, Karachi, Pakistan.
\email{faraz.a.zaidi@ieee.org}
\and
Rozenblat, C \at University of Lausanne, Institute of Geography and Sustainability, Lausanne, Switzerland.
\email{celine.rozenblat@unil.ch}
}

\date{Received: date / Accepted: date}

\maketitle

\begin{abstract}
One perspective to view the economic development of cities is through the presence of multinational firms; how subsidiaries of various organizations are set up throughout the globe and how cities are connected to each other through these networks of multinational firms. Analysis of these networks can reveal interesting economical and spatial trends, as well as help us understand the importance of cities in national and regional economic development. This paper aims to study networks of cities formed due to the linkages of multinational firms over a decade (from 2010 to 2019). More specifically we are interested in analyzing the growth and stability of various cities in terms of the connections they form with other cities over time. Our results can be summarized into two key findings: First, we ascertain the central position of several cities due to their economically stable connections; Second, we successfully identify cities that have evolved over the past decade as the presence of multinational firms has increased in these cities. 

\end{abstract}

\section{Introduction}\label{sec::introduction}



Multinational firms play an essential role in the economic development of cities. The presence of multinational firms ensures a prosperous ecosystem for a city as it positively contributes towards tax revenue, employment rate, economic activity, and improved gains for domestic firms \cite{keller2009multinational,melnyk2014impact,abor2008foreign}. Different cities around the world are connected via their linkages through these multinational firms, impacting economies at both domestic and international levels. A city where numerous multinational firms are established can strongly influence the region where it is situated. New York, for example, is a mega-city housing headquarters of several multinational firms. It is highly connected domestically as well as internationally. Economic changes in New York not only impacts the US but other world regions that are connected to, and dependent on the economic stability of New York.

Many researchers have studied the role of multinational firms in the development of cities  \cite{derudder03,derudder2011global,belderbos2020firms,iammarino2018international}. As new firms are established, or existing firms are acquired by multinationals, city to city links also change. This change can be studied over time to understand how urban and regional economies shape as a function of multinational firms and their links with other cities \cite{hussain19,rozenblat17}. Typically, we can classify the change as either \textit{stable} or \textit{changing}. Cities that maintain their connections with other cities over time can be called \textit{stable cities}, whereas cities whose linkages vary over time can be called \textit{changing cities}. Economic stability of a city can be directly associated with cities that are able to maintain economic ties with other cities. For a city that changes its connections, we can say that the city demonstrates economic growth (or contraction), as new linkages are established (or previous ones are broken) \cite{fritsch2004effects}.


In this paper, we study the networks of cities formed by multinational firms as they change from the year 2010 to 2019. We used a metric called topological overlap to determine stable and changing cities in these networks. Topological overlap is a temporal measure that determines the similarity between two networks by considering the changes in links between the nodes. Two key findings from this analysis reveal the role of multinational firms in the economic stability of cities as well as successfully identify cities that have shown economic development in the context of multinational firms.

The paper is organized as follows: We discuss related work in Section 2, followed by details regarding the dataset used in this study in Section 3. In Section 4, we discuss the methodology used to analyze the networks. We commit Section 5 to Results and Discussion and Section 6 to wrap up with Conclusion and Future research prospects. 

\section{Related Work} \label{sec::related}
The role of multinational firms in the context of networks of cities has been studied extensively in various contexts to understand their impact on global economies. Taylor and Derudder thoroughly assessed the overall connectivity of city networks using multinational firms\cite{Taylor2003WorldCN}. Derudder also explored the hierarchical tendencies and regional patterns formed within the city networks through links of multinational firms\cite{derudder03}. Joyez studied the network of French multinational firms and the global value chains to reveal the alignment between the two networks\cite{joyez19}. Silva et al. used wire transfers between different cities to show how the transfers are directly proportional to their dependence on each other's economy \cite{silva2020}. Using the links for Advanced Producer Service firms with their Initial Public Offerings clients from the Hong Kong stock exchange from 1999 to 2017, Pan et al. highlighted the role of Chinese cities in reshaping the global economy network \cite{pan2018}. Using service provisioned connections between Advanced Producer Service firms and their clients, they also showed dominant cities in the Chinese urban network \cite{pan2017mapping}.

Analysis of networks of cities can help understand various economic phenomena. Mahutga et al. analyzed the shift of power distribution using airline passengers data and its implication on the connectivity of cities \cite{mahutga10}. Blumenfeld-Lieberthal used the topology of a transportation network of various cities to show a positive correlation of economic activity, and high connectivity between two cities \cite{blumenfeld09}. Interdependence within the network of cities arising from global and local processes has been thoroughly examined by Sassen \cite{sassen07}. Sassen also analyzed the impact on the social structures emerging from these relationships and pointed out the importance of global cities \cite{sassen11}. Ducruet et al. analyzed cities through a maritime network to show the relevance of cities in maritime networks considering both network-specific and classical factors \cite{ducruet2018maritime}. Using a sea-port network of 91 countries, Munim and Schramm found a correlation between the quality of port infrastructure with higher economic growth for developing and developed countries, albeit a bit weaker for the latter \cite{munim2018impacts}. By studying the high-speed rail network, Diao found growth in fixed asset investments for cities with high-speed rail service in China \cite{diao2018does}. 

Network of cities also provide a dimension to rank cities based on different metrics. Cities were ranked based on diversity, centrality, and strength by Hussain et al. to identify established and changing cities \cite{hussain19}. In the context of climate change, Wang et al. ranked various cities using their economic performance and mitigation of climate changes based on Data Envelopment Analysis \cite{wang2020}. Wang also evaluated the performance of global cities based on Data Envelopment Analysis and Malmquist index taking into consideration various aspects such as economy, R\&D, etc. \cite{wang19}.

Pan et al.\cite{pan2018world} study the networks created from the global financing of Chinese firms using data from initial public offerings (IPO) on the Hong Kong stock exchange. The authors draw interesting conclusions based on economic activities to study city-to-city links. Similar to our dataset, we analyze all economic linkages (not just as a result of IPO) to understand how world city networks have shaped over the past decade. Zhao et al. \cite{zhao2020} study service firm networks and the resulting connections of cities to identify core and peripheral cities in China. Interesting observations about the provincial capitals is found and discussed in the context of developing the Chinese economy.  

Rozenblat \cite{rozenblat21} study the intercity and intracity networks of cities at meso-level and macro-level. This is to understand the complex interplay of spatial scale and power as different cities and regions demonstrate various strategies to develop and strengthen local economy and compete internationally through global expansion.

Temporal networks have traditionally proven to be helpful in studying change. Temporal networks were broadly studied by Holme and Saram{\"a}ki, where they defined different types of temporal networks and introduced various measures for the study of networks that change over time (equivalent to their static counterpart) \cite{holme12}. They also stated other existing temporal network models and ways to represent temporal data in the form of static graphs. Raimbault studied the temporal behavior of networks to observe the co-evolution of the French system of cities with the transportation network \cite{raimbault2021}. Bagan and Yamagata presented an approach to monitor land-cover dynamics of 50 global cities using their temporal Landsat data of 25 years \cite{bagan2014}. 

Many metrics have been used with various temporal networks in analyzing network properties. Zhao et al. used temporal centrality on a stock market based network as a portfolio selection tool \cite{zhao2018stock}. They emphasized the importance of temporal attributes of the network for real-world applications. Prado et. al determined the most relevant characters in a story by using eigenvector centrality over a temporal network of literary text \cite{prado2016temporal}. Cunha et al. analyzed contact duration and inter-contact times in vehicular networks to compare original and calibrated mobility traces of GPS-equipped taxis \cite{cunha2016communication}. Clauset and Eagle introduced topological overlap as a similarity measure between two networks (the two networks can be consecutive snapshots of the same network) \cite{clauset2012}. Buttner et al. adapted the metric for edge cases and used it on a real-world animal trade network \cite{buttner2016}. Navarro et al. used topological overlap to measure embeddedness of a link to show correlation with tie persistence.\cite{navarro2017}.

Utilizing the methodical information gained from the aforementioned studies, we used a network of cities connected via linkages of multinational firms. These linkages allow us to study stable and changing cities from the perspective of economic growth. We use a temporal metric to examine networks of cities from 2010 to 2019 and use this data to identify the top stable and changing cities over the time period. 

\section{Dataset}\label{sec::data}
For the study, we used networks of cities in the context of multinational firms for the years 2010 and 2019. A link in this network represents a connection between two cities for which there exist two firms belonging to the same group of companies with a financial tie. A financial tie means that a firm owns a share in the capital of another firm. A node in this network is not exactly a city but large urban areas labeled with the same criteria worldwide. We selected the top 3000 group of companies based on their annual turnover and their 800,000 direct and indirect subsidiaries accumulating over 1.3 million links for the years 2010 and 2019. This data is provided by Bureau van Dijk\footnote{Bureau van Dijk Electronic Publishing (\url{http://www.bvdep.com/)}}. Subsequent data cleansing was done as a collaboration between Universit\'{e} de Lausanne - CitaDyne research group and University of Paris - ERC GeoDiversity research group\footnote{ORBIS 2010, 2013: BvD - Universite de Lausanne (CitaDyne group) and University of Paris (ERC Geodiversity group).} to correct erroneous entries and fill in missing values. The network is similar to other datasets used by researchers \cite{alderson04,wall11,kraetke14} in that they all represent economic networks of cities. However,  our dataset is  comparatively larger. Another necessary difference in the network is the removal of the self-loop since we are primarily concerned with the linkages between the cities rather than within them. 

Figure 1 \cite{rozenblat17} demonstrates the construction of these networks. To show the construction, we use three groups of companies as an example. Figure 1a shows the three groups: Group 1, Group 2, and Group 3 colored Pink, Green, and Dark Blue respectively. They are present in three different cities: City A, City B and City C colored Yellow, Red, and Light Blue respectively. Figure 1b shows how the cities are connected through the branches of the above-mentioned companies spread across the cities. Figure 1c shows how a simple graph\footnote{A simple graph has only one edge between two vertices with no self-loop} is created by aggregating links between the cities. 

Mathematically, we can define the cities' network as:
\begin{equation}
	G' = (V',E')\label{eq:graph1}
\end{equation}
\begin{equation}
	G'' = (V'',E'')\label{eq:graph2}
\end{equation}

where \begin{math}G'\end{math} is the cities' network for the year 2010, comprising of nodes \begin{math}V'\end{math} and edges \begin{math}E'\end{math} and, \begin{math}G''\end{math} is the cities' network for the year 2019, comprising of nodes \begin{math}V''\end{math} and edges \begin{math}E''\end{math} such that \begin{math}V'  \subseteq V''\end{math}. 

Table \ref{table:networkstats} shows a summary of standard network metrics for the two networks. The networks are slightly different if we see their respective average path lengths and transitivity. We also see a relatively significant difference in the number of nodes, number of edges, the edge-node ratio, and the highest degree between the two networks. This is due to the growth of the network over the studied period, where new cities became part of the network, and current cities further consolidated their presence in the network by forming new links with other cities.

\begin{figure}[ht]
	\centering
	\includegraphics[width=0.9\linewidth]{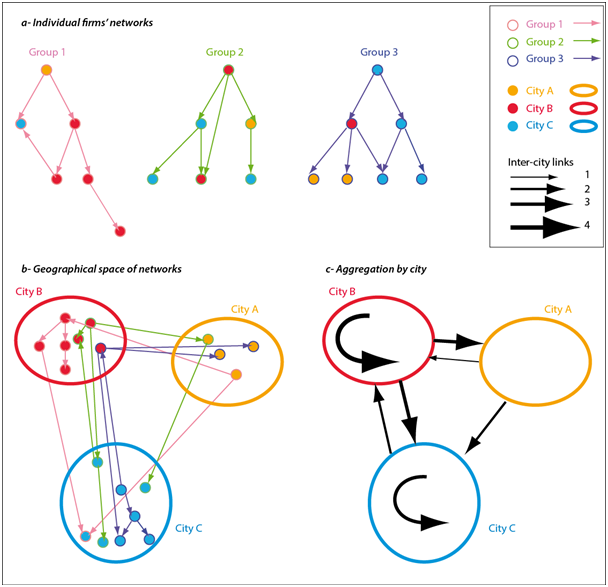}
	\caption{Economic Networks of Cities: (a) Hierarchical linking of Multinational Firms to their branches (b) Firms are Geographically Distributed and (c) Link aggregation to form connection between cities. Image Source: Originally constructed to explain how the multinational firms' networks were built \cite{rozenblat17}.}
	\label{fig::citynetwork}
\end{figure}

\begin{table}[hbt!]
	\centering
	\caption{Network Statistics: Comparison of the two networks. Structural similarity between the two networks is evident from global metrics.
	}
	\begin{tabular}{|l|c|c|}
		\hline \rule[-2ex]{0pt}{5.5ex} \textbf{Metric} & \textbf{2010} & \textbf{2019} \\ 
		\hline Nodes & 1210 & 1423 \\ 	
		\hline Edges & 32374 & 54048 \\ 
		\hline Edge-Node Ratio & 26.75 & 37.98 \\ 
		\hline Highest Degree & 737 & 873 \\ 
		\hline Average Path Length & 2.29 & 2.24 \\ 
		\hline Transitivity & 0.343 & 0.393 \\ 
		\hline 
	\end{tabular} 
	\label{table:networkstats}
\end{table}

\section{Methodology}\label{sec::identifying}
\begin{figure}[ht]
	\centering
	\includegraphics[width=0.75\linewidth]{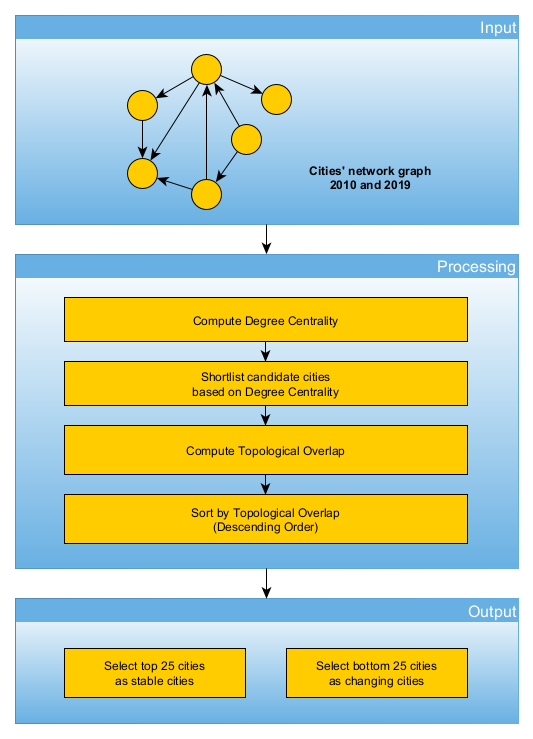}
	\caption{Methodology for identifying stable and changing cities in the network. Candidate cities are shortlisted using the degree centrality; topological overlap is then computed for these cities to find stable and changing cities. }
	\label{fig::methodology}
\end{figure}

Figure \ref{fig::methodology} outlines the methodology used to analyze the dataset. We first compute degree centrality for each city and use that as a criterion to shortlist important cities. Then, we calculate the topological overlap for the shortlisted cities and sort the results in descending order. Cities with high topological overlap are classified as stable cities, while those with low topological overlap are identified as the cities with the most topological change over the studied period.

Topological overlap \cite{clauset2012} is a similarity measure that we can use to define similarity between the connections of a city over the next interval. The topological overlap for node i between \begin{math}t_{m}\end{math} and  \begin{math}t_{m+1}\end{math} is defined as:

\begin{equation}
C_{i}(t_{m},t_{m+1})= \dfrac{\sum_{j} a_{i,j}(t_{m}) a_{i,j}(t_{m+1})}{\sqrt{\sum_{j} a_{i,j}(t_{m})\sum_{k} a_{i,k}(t_{m+1})}}\label{eq:3}
\end{equation}

subject to $\sum_{j} a_{i,j}(t_{m}) > 0$ and $\sum_{j} a_{i,j}(t_{m+1}) > 0$\par

Here, \begin{math}C_{i}\end{math} is the topological overlap of node $i$,   \begin{math}t_{m}\end{math} and \begin{math}t_{m+1}\end{math} are the time intervals for which topological overlap is computed. \begin{math}a_{i,j}(t_{m})\end{math} and \begin{math}a_{i,j}(t_{m+1})\end{math} are link's weight for weighted graphs (1 or 0 for unweighted graphs), corresponding to weighted presence or absence of a link between node i and node j, determined through adjacency matrix of the network for time \begin{math}t_{m}\end{math} and \begin{math}t_{m+1}\end{math} respectively.\\

Two important properties help us to understand how topological overlap works. The first property is that the topological overlap of a node $i$ is $0$ if there is no overlap between the links of node $i$ for the two time periods $t_{m}$ and $t_{m+1}$. Second, the topological overlap is greater than $0$ when there is some overlap between the two graphs at time intervals $t_{m}$ and $t_{m+1}$. We show the proofs of these two properties below:
	
\textbf{Property 1:} \begin{math} C_{i}(t_{m},t_{m+1}) = 0 \end{math}, when there is no overlap\newline

\textbf{Proof:} The numerator in equation \eqref{eq:3} is a summation of the product of two terms, $a_{i,j}(t_{m})$ and $ a_{i,j}(t_{m+1})$. If any of the two terms is zero, the respective term in the summation will be zero. If all terms in the summation are zero, $C_{i}(t_{m},t_{m+1})$ will be zero.\newline In case when there is no overlap between $t_{m}$ and $t_{m+1}$ for node $i$, then $a_{i,j}(t_{m}) \cdot a_{i,j}(t_{m+1}) $ will always be zero because, by definition\newline\par 
{\centering
	$a_{i,j}(t_{m}) = 0$\quad if $a_{i,j}(t_{m+1}) > 0 $,\newline $a_{i,j}(t_{m+1}) = 0$ \quad if $a_{i,j}(t_{m}) > 0$\newline\par
}
Hence, all the product terms in the summation will be zero and therefore, $C_{i}(t_{m},t_{m+1})$ will be zero.\newline
 
\textbf{Property 2:} \begin{math} C_{i}(t_{m},t_{m+1}) > 0 \end{math}, if there is overlap between graph $t_{m}$ and $t_{m+1}$\newline

\textbf{Proof:} In contrast to the statement in the above proof, if even a single term in the summation of the numerator of equation \eqref{eq:3} is greater than zero, $C_{i}(t_{m},t_{m+1})$ will not be zero.\newline In case when there is overlap between $G'$ and $G''$ for node $i$ and node $j$, then $a_{i,j}(t_{m}) \cdot a_{i,j}(t_{m+1}) \forall j \in V''$ will be greater than zero because, by definition\newline\par 
{\centering
	$a_{i,j}(t_{m}) >  0$\quad if $a_{i,j}(t_{m+1}) > 0$\newline\par
}
Hence, at least one product term in the summation will be greater than 0 and therefore, $C_{i}(t_{m},t_{m+1})$ will be greater than zero.\newline
	 
Topological overlap is generally used with binary links. The output interpretation is relatively straightforward; if there is no change in the link for a city over the next interval, topological overlap will be 1; for a complete change of links, topological overlap will be 0. Partial changes will then lie between 0 and 1. Weighted topological overlap has also been used in the past to study networks where the links have a certain strength \cite{mumford2010}. The downside of this approach is that the values are no longer constrained between 0 and 1, making it irrelevant for certain comparisons but still useful in the absolute sense. 

In this study, we have used weighted links to compute topological overlap\cite{teneto2020}. The objective for this selection was to capture the effect of high connectivity between the cities in the topological overlap value. The weight of a link is determined by the number of financial ties between two cities. Furthermore, we used directed links to compute topological overlap taking into account the parent/subsidiary relationship between two cities.

To illustrate the point, take two cities with different connectivity patterns that have maintained all their links over the next interval. In the case of binary links, since the output is constrained between 0 and 1, the topological overlap will be 1 for both cities. However, we will see that cities with higher connectivity will have a proportionally higher topological overlap in the weighted case.

\begin{figure}[ht]
	\centering
	\includegraphics[width=1\linewidth]{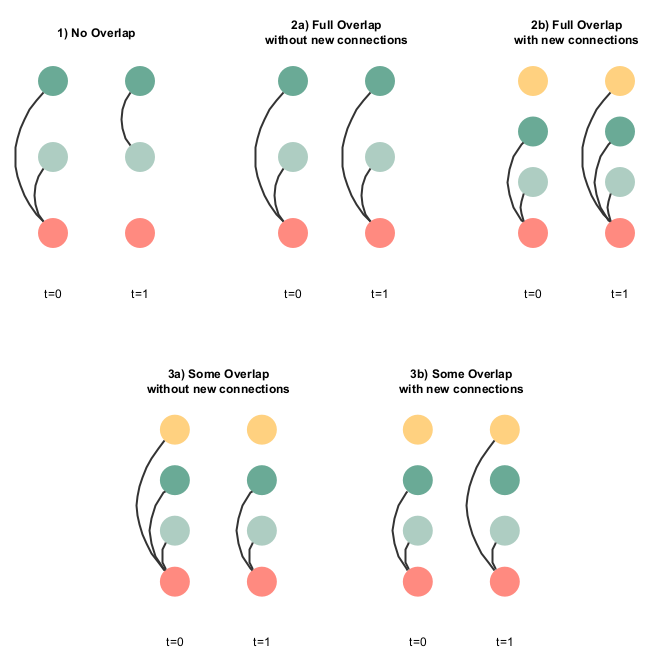}
	\caption{Various cases for evolution of city's connection over time}
	\label{fig::overlapcases}
\end{figure}



The connections of a city can evolve over time in the following manner (see Figure \ref{fig::overlapcases}):
\begin{itemize}
	\item No connections are retained
	\item All connections are retained (a) with no new connections and (b) with new connections
	\item Some connections are retained (a) with no new connections and (b) with new connections
\end{itemize}

To get meaningful insights from the topological overlap, we were required to shortlist candidate cities. This step was not necessary for identifying stable cities as weighted links amplify topological overlap values for them. However, important changing cities could easily be mistaken due to the behavior of topological overlap for maximum change, which will return zero. For example, take two cities with degree centralities of 1 (City A) and 100 (City B), respectively. Over the next time interval, City A does not retain any connection, and City B retains 1 connection out of 100. The topological overlap for City B, in this case, will be more significant than City A, which is a misleading result. We used degree centrality as a centrality measure to shortlist the top 250 cities filtering out all cities with low connectivity.

Before proceeding towards results and discussion, it would be interesting to see whether topological overlap offers a different perspective to determine changing cities or not. A simpler way to observe changing cities would be to observe the change in the degree centrality of the cities from 2010 to 2019. Table \ref{tbl::to_dc_rank} compares the rank of top changing cities identified through degree centrality and topological overlap. Two lists are presented (see Table \ref{tbl::to_dc_rank}), one where top 25 cities based on Topological overlap (TO-Rank) are listed along with their relative rank in terms of degree centrality (DC-Rank); the second list is where the top 25 cities with respect to change in Degree centrality (DC-Rank) from 2010 to 2019, and their respective ranking in terms of topological overlap (TO-Rank) are identified. Considering the top 25 cities that have changed over the studied period based on weighted degree centrality, we can clearly see that cities from the US and China dominate the list. Beijing, Shanghai, Hong Kong and Shenzen, all represent the strong developing economy of China around these central cities where the presence of multinational firms has increased their local as well as global ties. On the other hand, cities such as Washington, Chicago, Los Angeles, Dallas, Philadelphia, New York, Houston, San Francisco, Atlanta, Denver, Wilmington, Nashville, Boston, Miami and Detroit represent the strong domination of multinational firms headquartered in the US in the world economy. In terms of topological overlap, top cities identified as changing cities are again dominated by cities from the US and China. Cities from China include Urumqi, Zhengzhou, Wenzhou, Jiujiang, Qingdao, Jinan, Hefei, Ganzhou, Chongqing and cities from the US include Lafayette, Grrenville, Shreveport, Fort Myers, Little Rock, Dayton, Knoxville and Flint. These cities are different than what we see in the list based on degree centrality. This is because weighted degree centrality is a good measure to capture absolute change in the number of connections whereas topological overlap focuses on the change in terms of the structure of connections and thus reveals a different set of cities altogether. A detailed discussion and our findings on the cities that stand out is given in section \ref{sec::results} below.


\begin{table}[]
	\centering
	\caption{List of Top Changing Cities Ranked using Topological Overlap and Degree Centrality for the year 2010 and 2019. TO-Rank represents the ranking of cities based on Topological Overlap and DC-Rank represents the ranking of cities based on Weighted Degree Centrality.}
	\begin{tabular}{|c|c|c|c|c|c|}
		\hline
		\textbf{City} & \textbf{TO-Rank} & \textbf{DC-Rank} & \textbf{City} & \textbf{TO-Rank} & \textbf{DC-Rank} \\ \hline
		Doha        & 1  & 147 & Beijing       & 192 & 1  \\ \hline
		Urumqi      & 2  & 112 & Tokyo         & 237 & 2  \\ \hline
		Zhengzhou   & 3  & 72  & Washington DC & 191 & 3  \\ \hline
		Wenzhou     & 4  & 95  & Chicago       & 221 & 4  \\ \hline
		Lafayette   & 5  & 140 & Los Angeles   & 210 & 5  \\ \hline
		Greenville  & 6  & 156 & Dallas        & 201 & 6  \\ \hline
		Shreveport  & 7  & 142 & Philadelphia  & 222 & 7  \\ \hline
		Jiujiang    & 8  & 164 & New York      & 243 & 8  \\ \hline
		Abu Dhabi   & 9  & 141 & London        & 239 & 9  \\ \hline
		Fort Myers  & 10 & 83  & Houston       & 190 & 10 \\ \hline
		Qingdao     & 11 & 114 & Shanghai      & 184 & 11 \\ \hline
		Jinan       & 12 & 64  & San Francisco & 225 & 12 \\ \hline
		Hefei       & 13 & 99  & Atlanta       & 188 & 13 \\ \hline
		Ganzhou     & 14 & 93  & Dublin        & 194 & 14 \\ \hline
		Little Rock & 15 & 130 & Singapore     & 226 & 15 \\ \hline
		Boise       & 16 & 135 & Denver        & 148 & 16 \\ \hline
		Dayton      & 17 & 175 & Wilmington    & 213 & 17 \\ \hline
		Cork        & 18 & 107 & Grand Cayman  & 244 & 18 \\ \hline
		Limassol    & 19 & 146 & Nashville     & 173 & 19 \\ \hline
		Colombia    & 20 & 133 & Hong Kong     & 235 & 20 \\ \hline
		Chongqing   & 21 & 76  & Shenzen       & 122 & 21 \\ \hline
		Knoxville   & 22 & 115 & Boston        & 234 & 22 \\ \hline
		Flint       & 23 & 118 & Luxembourg    & 224 & 23 \\ \hline
		Changsha    & 24 & 101 & Miami         & 162 & 24 \\ \hline
		Madison     & 25 & 132 & Detroit       & 153 & 25 \\ \hline
	\end{tabular}
	\label{tbl::to_dc_rank}
\end{table}

\section{Results and Discussion}\label{sec::results}
Following the steps mentioned earlier, we obtain a list of cities sorted by decreasing topological overlap. At the top of this list are the cities that retain their connections from 2010 to 2019; We call these cities, \textit{stable}. In contrast, at the bottom of the list are cities that changed most of their connections from 2010 to 2019; and these are called \textit{changing} cities.

\begin{figure}[ht]
	\makebox[\textwidth]{\includegraphics[clip,trim=8 0 40 0,width=1\textwidth]{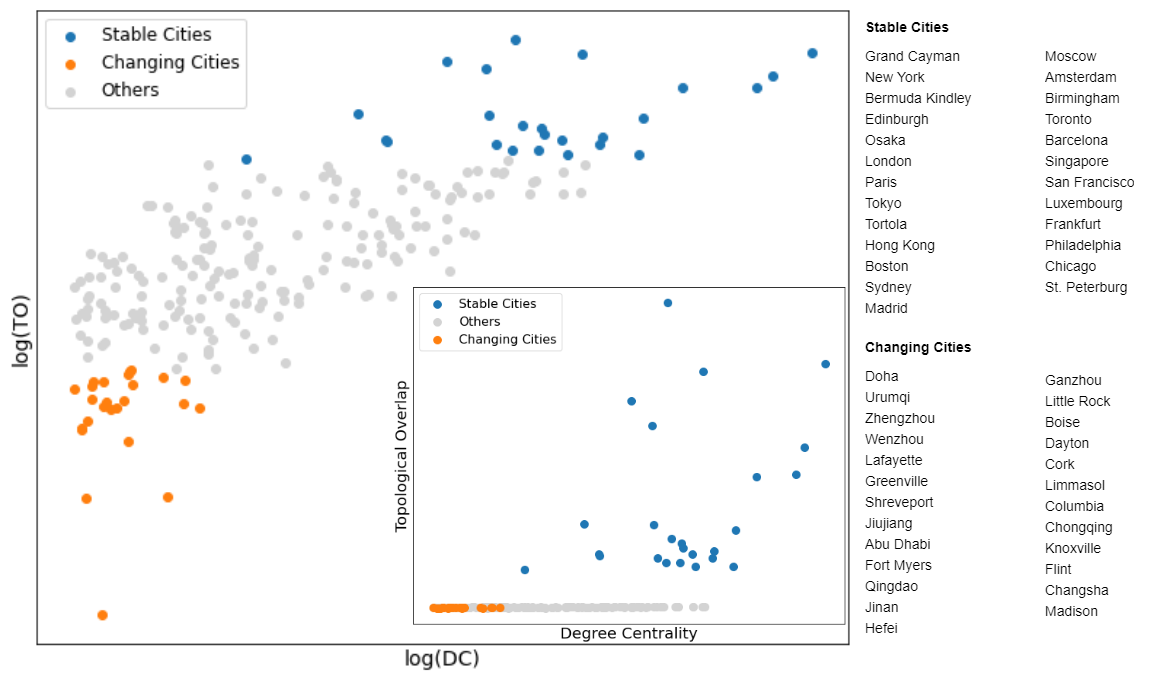}}
	\caption{The figure shows the selection of stable(blue) and changing cities(orange) from the top 250 cities shortlisted using degree centrality. log(DC) and log(TO) represent log of degree centrality and topological overlap, respectively. Inner figure: Plot of linear degree centrality vs. topological overlap; changing cities are not distinguished from other cities due to low degree centrality and topological overlap.}
	\label{fig::dc_to}
\end{figure}

\subsection{Discussion for stable cities}
Table \ref{tbl::topStable} shows the list of stable cities identified through topological overlap. At first glance, cities like New York, London, and Paris emerge; The presence of these cities in this list reaffirms results from previous studies about their economic importance in the world \cite{sassen01,zukin2005,hussain19}. These cities are global hubs\footnote{Global hubs are centers for many multinational firms} with very high degree centralities, bringing economic stability to not only their countries but to their regions as well.

Apart from London and Paris, other cities of the Western European region also make this list. Amongst theses is Edinburgh, which is at the top of the list in the region (ahead of London and Paris), Madrid, Amsterdam, Birmingham, Luxembourg, and Frankfurt. Edinburgh is considered amongst one of the strongest economies in the region, with a strong financial services industry as its backbone. Birmingham makes it to the list as a result of its manufacturing industry and its strong inclination towards entrepreneurship. Edinburgh and Birmingham's high economic outputs contribute to maintaining a steady economy for the UK. Madrid is the financial and industrial center for Spain. Despite being an industrial hub, the services sector is the major contributor to the Spanish economy. For the Netherlands, Amsterdam is the financial hub housing major banking corporations. Being a port city, Amsterdam also contributes to the country's economy via national and international trade. The city of Luxembourg is considered for its financial services, with the financial sector being the largest contributor to its country's economy. Lastly, Frankfurt, one of Germany's financial centers, has contributed over the time period studied to the German economy via its banking sector consisting of major banks. Frankfurt also has a strong aviation presence in the world. Both these industries help Frankfurt provide economic stability to Germany.

\begin{figure}[ht]
\makebox[\textwidth]{\includegraphics[clip,trim=40 0 40 0,width=1\textwidth]{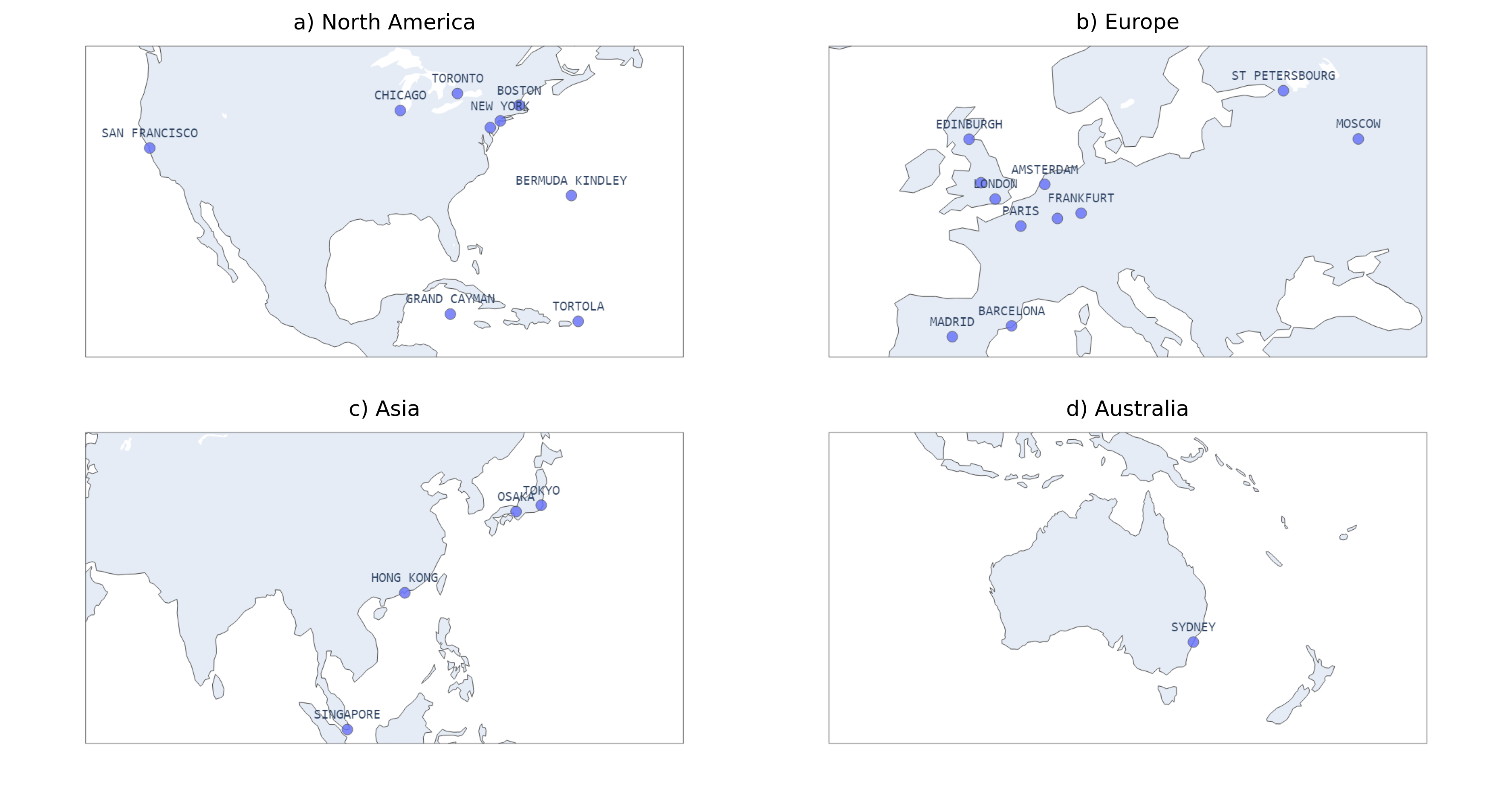}}
	\caption{Top stable cities identified by region: a) North America b) Europe c) Asia and d) Australia}
	\label{fig::stable}
\end{figure}
In the Eastern European region, the top two cities of Russia, Moscow and St. Petersburg, are identified amongst the stable cities in the list. Moscow is the financial center for Russia and is home to many of Russia's largest firms, including its major tech companies. St. Petersburg is a port city and is pivotal to the Russian economy. It also has a stable automotive industry. Both the cities have been vital for Russia's strong economy over the years.

\begin{table}[ht]
	\centering
	\caption{List of Top 25 Stable Cities Based on Topological Overlap for 2010 and 2019}
	\begin{scriptsize}
		\begin{tabular}{|l|c|c|c|l|c|c|c|}
			\hline \textbf{Rank} & \textbf{City} & \textbf{Country} & \textbf{Region} & \textbf{Rank} & \textbf{City} & \textbf{Country} & \textbf{Region} \\   
			\hline 
			1. & Grand Cayman		& Cayman Islands& Carribean			& 14. & Moscow			& Russia		& Eastern Europe	\\	 
			2. & New York         	& USA	 		& North America		& 15. & Amsterdam	    & Netherlands	& Western Europe	\\	
			3. & Bermuda Kindley   	& Bermuda 		& Caribbean			& 16. & Birmingham	    & UK			& Western Europe	\\	
			4. & Edinburgh         	& UK		 	& Western Europe	& 17. & Toronto	 	   	& Canada		& North America		\\	
			5. & Osaka         		& Japan 		& Eastern Asia		& 18. & Barcelona		& Spain			& Western Europe	\\	
			6. & London         	& UK		 	& Western Europe	& 19. & Singapore	    & Singapore		& Eastern Asia		\\	
			7. & Paris         		& France 		& Western Europe	& 20. & San Francisco	& USA			& North America		\\	
			8. & Tokyo	         	& Japan 		& Eastern Asia		& 21. & Luxembourg		& Luxembourg	& Western Europe	\\	
			9. & Tortola		   	& BVI	 		& Caribbean			& 22. & Frankfurt	    & Germany		& Western Europe	\\	
			10. & Hong Kong        	& Hong Kong 	& Eastern Asia		& 23. & Philadelphia	& USA			& North America		\\	
			11. & Boston         	& USA 			& North America		& 24. & Chicago		    & USA			& North America		\\	
			12. & Sydney         	& Australia 	& Australia/Oceania	& 25. & St. Petersburg	& Russia		& Eastern Europe	\\	
			13. & Madrid         	& Spain 		& Western Europe	& 	  &	  			    & 	   			& 			 		\\	
			
			\hline
		\end{tabular} 
	\end{scriptsize}
	\label{tbl::topStable}
\end{table}

In the North American region, Boston, Chicago, San Francisco, and Philadelphia from the USA also make it into the list, along with Toronto from Canada. Boston has one of the largest economies in the US and houses numerous tech companies, a large number of these geared towards biotechnology. This is primarily influenced by the schools and research centers in and around Boston attracting investment. On the other hand, Chicago is one of the most important financial centers of the world. Chicago also has a strong food and retail industry with major firms headquartered in the city. San Francisco is another financial center in the US and has a growing technology industry. Philadelphia also has a strong biotechnology industry and contributes to the economy via trade as a port city. Toronto has been a financial hub for Canada for many years. Due to its start-up ecosystem, it has also grown as one of the largest cities to host the tech industry in the North American region. Since all the industries mentioned above have remained relevant throughout the past decade, these cities have experienced stability in their linkages with other cities and thus feature in the list of most stable cities. 

Three cities from the Caribbean region, namely Grand Cayman, Bermuda Kindley, and Tortola, are also listed as stable cities. Financial services provided by these cities are responsible for a large part of their economy. These cities are home to many offshore companies attracted by their lucrative tax policies. 

Sydney is the only stable city identified in the Australia/Oceania region. It is the leading financial center in the region, and at the national level, most of the banks are based out of Sydney. Sydney also has the largest manufacturing industry in Australia. We can also see three important cities of the Eastern Asia region in the list: Tokyo and Osaka, from Japan, and Singapore. These cities are again important financial centers of the world. Japan also has one of the largest industry for electronics globally, which has remained steady over the last decade, and so has Japan's economy. Major multinational electronics firms have their headquarters situated in Tokyo and Osaka. Similarly, Singapore holds a significant share in the global semiconductors market and has been important for its stable economy.

Stable cities have continued to maintain their existing connections to gain economic stability. A majority of these cities are either financial centers or providers of financial services worldwide. Some cities are manufacturing giants, while others have relied on their strong technological foundations to maintain stability. 

\begin{table}[ht]
	\centering
	\caption{List of Top 25 Changing Cities Based on Topological Overlap for 2010 and 2019}
	\begin{scriptsize}
		\begin{tabular}{|l|c|c|c|l|c|c|c|}
			\hline \textbf{Rank} & \textbf{City} & \textbf{Country} & \textbf{Region} & \textbf{Rank} & \textbf{City} & \textbf{Country} & \textbf{Region} \\   
			\hline 
			1. & Doha				& Qatar		& Western Asia - Gulf	& 14. & Ganzhou         	& China 	& Eastern Asia	\\	 
			2. & Urumqi         	& China	 	& Eastern Asia			& 15. & Little Rock	    & USA			& North America	\\	
			3. & Zhengzhou   		& China 	& Eastern Asia			& 16. & Boise		    & USA			& North America	\\	
			4. & Wenzhou         	& China		& Eastern Asia			& 17. & Dayton	 	   	& USA			& North America		\\	
			5. & Lafayette         	& USA	 	& North America			& 18. & Cork			& UK			& Western Europe	\\	
			6. & Greenville        	& USA		& North America			& 19. & Limassol	    & Cyprus		& Western Europe	\\	
			7. & Shreveport      	& USA	 	& North America			& 20. & Columbia		& USA			& North America		\\	
			8. & Jiujiang        	& China 	& Eastern Asia			& 21. & Chongqing		& China			& Eastern Asia	\\	
			9. & Abu Dhabi		   	& UAE	 	& Western Asia - Gulf	& 22. & Knoxville	    & USA			& North America		\\	
			10. & Fort Myers       	& USA		& North America			& 23. & Flint			& USA			& North America		\\	
			11. & Qingdao         	& China 	& Eastern Asia			& 24. & Changsha	    & China			& Eastern Asia		\\	
			12. & Jinan         	& China 	& Eastern Asia			& 25. & Madison		    & USA			& North America	\\	
			13. & Hefei         	& China 	& Eastern Asia			& 	  &	  			    & 	   			& 			 		\\	
			
			\hline
		\end{tabular} 
	\end{scriptsize}
	\label{tbl::topChanging}
\end{table}

\subsection{Discussion for changing cities}

Table \ref{tbl::topChanging} shows the list of cities where significant change can be observed between the year 2010 and 2019. To reiterate, we are labeling a city as \textit{changed} if its connections are not similar over the studied time periods. It can also be interpreted as if some, or all the links from the previous timestamp are not retained thus resulting in being identified as changing cities. 

Doha (Qatar), is identified as the city with the most significant change on this list. Doha's economy is based on oil and gas industries, with many firms headquartered in the city. It is also expected to host the 2022 FIFA World Cup as Qatar spent heavily to develop infrastructure before the event. The infrastructure changes include expanding airports and constructing new road and rail networks. Doha's most substantial ties are formed within the Middle East with Bahrain and UAE. Outside the region, Doha is also strongly connected with London and with tax havens, Grand Cayman, and Kwajalein. Abu Dhabi is another Middle Eastern city on this list. To restrict its economic reliance on oil, UAE has invested heavily to diversify its economy.

Many cities where change can be observed have been identified in China, with ten cities making it to the list. These cities have grown exponentially, with only a handful of connections in 2010 but hundreds of links by 2019. A majority of the links for these cities are local, bringing economic stability to different regions in the country. There also have been initiatives to develop infrastructure, attracting foreign investments to connect them with international cities. Geographically, the economic decentralization strategy also comes into play since the cities are not concentrated in a single region. Decentralization means less dependence on a few critical cities for economic prosperity, providing better resilience to China. In this case, if an important city suffers contraction, the economy will not crash for the whole country and the damage will be limited to the respective region. Moreover, it gives better chances to the affected city to recover quickly through the support of other cities. Apart from this economic decentralization, geographic decentralization can also be observed with cities emerging from different regions.

 \begin{figure}[ht]
 	\makebox[\textwidth]{\includegraphics[clip,trim=40 0 40 0,width=1\textwidth]{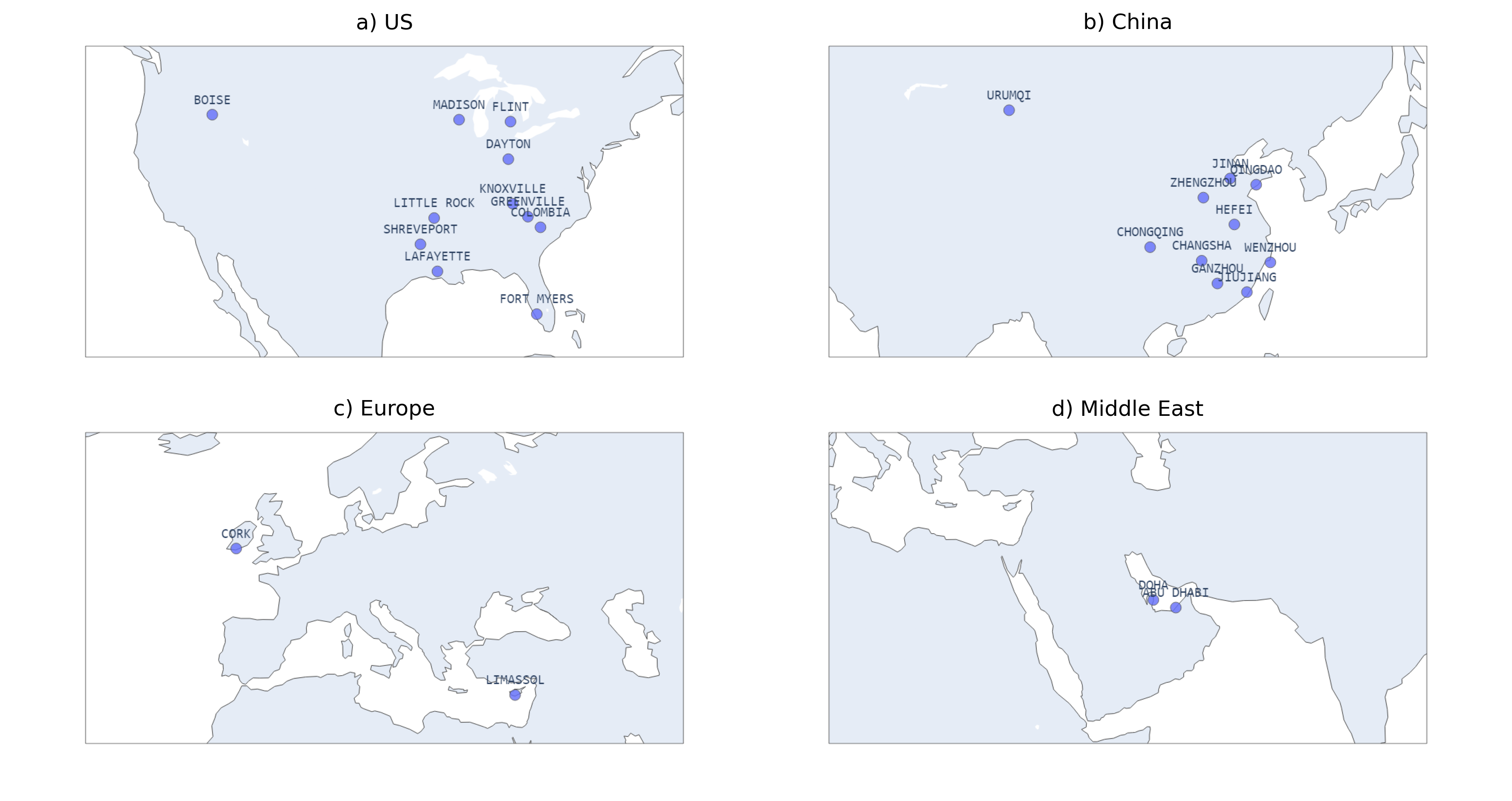}}
 	\caption{Top changing cities identified by region: a) US b) China c) Europe and d) Middle-East}
 	\label{fig::changing}
 \end{figure}
 
For example Urumqi from the province of Xinjiang in the far northwestern region of China is China's attempt to access markets from Central Asia. Urumqi has been identified as the city with the highest change in China. To promote foreign trade, the construction of a free trade zone was started in 2015 in Urumqi. It also hosted the 1st China-Eurasia Expo in 2011, an upgrade from its regional trade fair. The expo was intended to encourage trade not only from Central Asia but also from Europe and other Asian regions. To improve its connectivity, both domestically and internationally, various modes of transportation have also been upgraded for the city.

The city of Chongqing also appears in the list which has evolved as the commercial capital of southwestern China. The city is one of four municipalities under direct administration of central government (equivalent to a province). Manufacturing industry is the backbone of its economy and it is one of the largest vehicle manufacturers in China. The region is also rich in natural resources and agriculture. Recent developments in infrastructure has led to the arrival of many multinational corporations. It was identified as one of the top-performing economies in the world by Brookings Institution in 2018 \cite{brookings2018}.

Changsha is the largest city in the Hunan province from central China and is the 17th most populous city in China. It is located on the banks of Xiang River, which is 30 miles south of the Dongting Lake. This gives the city access to water ways all across the southern and southwestern parts of Hunan. The city is a major trade center with rice, cotton, timber and livestock being traded here. It is also an important railway connection from Hankou to Guangzhou and as a result,  has attracted handsome investments and many multinational firms to make it into the list. Rice milling, oil extraction, tobacco, textile and agricultural chemicals and fertilizers are some of the key industries boosting the city's economy.

Apart from these three inland cities (Urumqi,Chongqing and Changsha), the other cities that appear in the list are either port cities, or cities from provinces neighboring port cities. Starting from the northeastern province of Shandong, two cities Qingdao and Jinan appear in the list. Qingdao, a port city is famous for its brewery industry and manufacturing white goods (large home appliances) and electronics. Since it is a vital port, it has attracted a lot of foreign investments in the past decade. Being a free trade zone, the presence of industrial zones has also helped its economic growth. Another city from the Shandong province is Jinan which is home to the top universities in the country, and access to academic talent has been pivotal for attracting foreign investments. Also, Jinan's Innovation Zone has the highest economic growth in the country \cite{zheng19}. Automobiles, information technology, and materials are the key industries for Jinan.

Wenzhou from the province of Zhejiang is another port city from southeastern part of China. The city is known for low-power electrical appliances and local firms have collaborated with major players internationally. It is home to many privately owned business involved in low-end manufacturing. The export of these goods are responsible for significant boost to China's economy. The Wenzhou Economic and Technological Development Zone has also been established which also offers high connectivity to other cities along with lenient economic policies.

Zhengzhou is the capital city of Henan province and is also known as the 'iPhone City'. It is the most prominent smartphone manufacturing site globally. Zhengzhou also has an Airport Economy Zone, started in 2013, allowing businesses to connect with other firms quickly and smoothly. Other economic and industrial zones were also established in the city to promote industrial growth in IT, pharmaceutical and materials sectors. Zhengzhou is also pivotal from logistics perspective as it is also one of the transport hubs in China. The city lies in the northern part of Henan, where the railway line accesses the Longhai line which connects the east to west. The province itself has an agricultural economy with cotton, silk, tobacco and vegetable oils as its primary produce.

Hefei is the largest city of the province of Anhui and like Jinan, an academic center with numerous universities and research centers present in the city. Hefei's economy is based on its manufacturing industry, with the automobile, information technology, materials, and chemicals being the key industries. Hefei was amongst the fastest-growing metropolitan economy in the world in 2014  \cite{brookings2014}. 

Jiujiang is another port city on this list and the only international trade port city from the province. It has attracted sizable foreign investments, boosting its economy significantly. Apart from its strong transport network, Jiujiang is supported by its automotive, energy, materials, and electronic industries within the Economic and Technological Development Zone.

Ganzhou, which was poverty-ridden, has seen significant development over the past decade. A reliable high-speed rail network has been set up, connecting it to its neighboring province, making it an important strategic location for the One Belt One Road plan. Lenient tax policies have been introduced to encourage the growth of industries in the city.

Chinese cities have focused on industries that have been relevant in the past decade. These industries primarily belong to technology, automotive, and manufacturing sectors \cite{zhang2021study,techakanjanakit2012,wubbeke2016made}. They have invested in infrastructure and introduced favorable policies, and in return, they have experienced positive economic growth. Many of the changing cities have ports\footnote{Ports are either seaports or inland ports; inland ports are similar to seaports but with smaller capacity and use inland waterways such as rivers instead of the sea.}, facilitating ease of trade. Interestingly, their focus has not only been on port cities but also on the landlocked cities such as Urumqi, which is vital for trade with Central Asia and Europe \cite{bird2020belt,wen2019impacts}.

Most cities with change have been identified from the USA. A majority of these cities are in the eastern states of the USA, except for Boise (Idaho), situated on the western coast.  
Lafayette is identified as the city with the biggest change for the USA. Major oil, retail, and IT firms are situated in the city. Oil and gas have been the key industry for Lafayette, but it has been on the decline as of late. Other than Lafayette, Shreveport is another city from Louisiana that has been identified for its change. Port of Shreveport is a designated foreign trade zone, and with immediate access to barge, motor freight, rail, and air transport facilities, it is lucrative for businesses to invest in the city. Shreveport is also growing as an IT industry and is part of the technology cluster forming along the I20 corridor. Fort Myers and Little Rock also have foreign trade zones. Like Shreveport, Little Rock also provides inter-modal logistic facilities for businesses. Greenville has evolved from a textile-based economy to a diverse economy based on automotive, energy, and manufacturing sectors. Many multinational companies have a presence in Columbia from various sectors such as manufacturing, health care, information technology, and energy. To increase investments, Dayton introduced Opportunity Zones within the city which provides substantial tax incentives to investors. Boise has been home to a few big technology companies, but recently, there has been a surge of tech start-ups in the city. As former technological hubs become competitive, entrepreneurs have approached Boise, contributing to its growth.
Madison is also growing steadily with the boom in technology companies, inclined towards biotechnology, influenced by academia's collaboration with local businesses. Knoxville has shown growth due to its energy and retail sector.
Despite the declining automotive industry and water, crisis \cite{pieper2017flint} in the city, Flint continues to maintain ties both locally and internationally, suggesting its growth.

The economic growth of changing cities in the US can be attributed to its entrepreneurial culture \cite{hafer2013entrepreneurship}. Opportunities in the technology sector have attracted many entrepreneurs to start new businesses, assisted by the collaboration of academia to provide a skilled workforce. Besides the technology industry, other sectors have also contributed to the economic growth of these cities. As a result, the connections of these cities have grown drastically. Like the Chinese cities, these changing cities have most of their links within the country. 

Two European cities were identified among the changing cities in the list. Cork is one of the port cities of Ireland. Over the past decade, Cork has evolved as an IT hub with major technological firms stationed there. Pharmaceutical companies have also invested significantly in the city. Limassol is the port city of Cyprus and is an essential source of revenue for the country. Limassol is experiencing growth in the economy as low corporate tax and port availability attracts investments from multinational firms.

Cities with the most change have been identified mainly in Asia and North America, many of them found in China and the US, respectively. Their connections are primarily local, providing stability to their regions. These cities have focused mainly on technology and manufacturing for their economic growth. The Middle East has focused on diversified economic growth and reduced their reliance on oil and gas industry. A couple of European cities also indicated growth in their connections, while cities from other regions showed no significant change in their connections.

\subsection{Summary of Results and Discussion}
Based on the discussion, an analysis of the topological overlap allowed us to reach some critical conclusions. Firstly, the stable cities identified in our study align with previous studies related to important cities of the world. These include New York, London, Paris, and Tokyo, which are home to many multinational firms. Secondly, the identified changing cities in China hinted towards a decentralized economic strategy as almost all the identified cities belonged to different provinces. Most of these Chinese cities have ports (either seaport or inland port) which are vital for trade \cite{ducruet10}. China's inclination towards international trade for economic growth is also highlighted by the development of the landlocked city of Urumqi, which is pivotal for its Belt and Road initiative. Lastly, the changing cities identified in the US showed less inclination towards international trade as only a few of them are port cities. Furthermore, a majority of their connections were found within the US with the highly connected Interstate Highway System facilitating domestic trade between cities \cite{jaworski2018interstate}. These factors suggest the US' self-dependence on the economic growth of its changing cities.

From a macroscopic point of view, analyzing the data set makes a strong case of a relational economy \cite{bathelt11} where interactions among global companies situated geographically around the world play an important role in the economic development of not just cities, but regions and even countries to some extent. The complex interconnected systems, inter-dependencies,  global exchanges, intertwined economies and, local and national policies all play a vital role in shaping the economic landscape of the world we live in. This study provides evidence that integrated relational economies result in creating and adding value to the modern successes of fast growing cities and countries.

\section{Conclusions and Future Work}\label{sec::conclusions}
In this paper, we studied the network of cities formed through multinational companies and their subsidiaries for 2010 and 2019. We used topological overlap as a metric to identify stable and changing cities in this network. The stable cities identified in this paper further strengthen previous studies on the importance of cities like New York, London, Paris, and Tokyo\cite{hussain19,king2015global}. Stable cities are home to many multinational firms, which contribute to the economic stability of these cities. Analyzing the list of changing cities, it is clear that most of them are in the US or China. Changing cities in China are not concentrated in a single region implying decentralization strategy for economic growth. Most of these cities are port cities, indicating a higher focus on international trade for economic growth. On the other hand, US cities showed more intent towards domestic trade as these cities are densely connected within the country, and only a few of them are port cities.\par
Coincidentally, all cities identified in the changing cities indicated growth. Identifying declining cities can also provide interesting insights and can be pursued in the future. This study has considered just two intervals separated by a decade to identify important cities. Considering intermediate intervals, we can get a deeper understanding of the development of cities. As it is unnecessary for the change to be  monotonous in each instance, we may observe swings in the growth over the studied time intervals. We may also observe cities exhibiting different growth rates and ending up with the same strength of linkages in the end. Further research in this direction can aid in making informed decisions and devising better policies to promote economic growth resulting in urban and regional development.

\section{Conflict of Interest Statement}

All authors declare that they have no conflicts of interest. 

\section{Data Availability Statement}

The data used in this study was made available by Bureau van Dijk (Bureau van Dijk Electronic Publishing \url{http://www.bvdep.com/}). The data cannot be shared publicly due to certain licensing restrictions.

\section{Acknowledgments}

The authors would like to thank the team members of CitaDyne research group at Universit\'{e} de Lausanne, Switzerland and ERC GeoDiversity research group at University of Paris, France to correct erroneous entries and fill in missing data. The authors would also like to thank the anonymous reviewers for their valuable feedback and comments that enabled us to improve the quality of this manuscript by adding insightful discussions. 

\bibliographystyle{plain}
\bibliography{visu}

\end{document}